\pdfoutput=1 % only if pdf/png/jpg images are used
\documentclass{JINST}

\title{Measurements and optimization of the light
  yield of a TeO$_2$ crystal}
\newcommand{\TEO}{$\mathrm{TeO}_2$}

\author{F.~Bellini$^{a,b}$,
  L.~Cardani$^{a,b}$,
  N.~Casali$^{c,d}$,
  I.~Dafinei$^b$,
  M.~Marafini$^d$,
  S.~Morganti$^b$,
  F.~Orio$^b$,
  D.~Pinci$^b$\thanks{Corresponding author},
  G.~Piperno$^{a,b}$,
  D.~Santone$^a$, 
  C.~Tomei$^b$ and
  M.~Vignati$^b$ \\
\llap{$^a$}Dipartimento di Fisica, Sapienza Universit\`a di Roma, I-00185, Italy \\
\llap{$^b$}Istituto Nazionale di Fisica Nucleare, Sezione di Roma, I-00185, Italy\\
\llap{$^c$}Dipartimento di Scienze Fisiche e Chimiche, 
Universit\'a degli Studi dell'Aquila, I-67100, Coppito (AQ), Italy\\
\llap{$^d$}Istituto Nazionale di Fisica Nucleare, 
Laboratori Nazionali del Gran Sasso, I-67010, Assergi (AQ), Italy\\
\llap{$^e$}Museo Storico della Fisica 
e Centro Studi e Ricerche "Enrico Fermi",
Piazza del Viminale 1, Roma, Italy\\

E-mail: \email{davide.pinci@roma1.infn.it}}

\abstract{Bolometers have proven to be good instruments to search for rare processes because of their 
excellent energy resolution and their extremely low intrinsic background.
In this kind of detectors, the capability of discriminating $\alpha$ particles from electrons 
represents an important aspect for the background reduction.
One possibility for obtaining such a discrimination is provided by the detection of the \v{C}erenkov 
light which, at the low energies of the natural radioactivity, is only emitted by electrons.
This paper describes the method developed to evaluate the amount of light 
produced by a crystal of {\TEO} when hit by a 511 keV photon.
The experimental measurements 
and the results of a detailed simulation of the crystal and the readout system
are shown and compared. \\
A light yield of about 52~\v{C}erenkov photons per deposited MeV was measured.
The effect of wrapping the crystal with a PTFE layer, with the aim 
of maximizing the light collection, is also presented.}


\begin{document}

\section{Introduction}

Tellurium dioxide ({\TEO}) crystals have proven to be superb bolometers for the 
search of neutrino-less double beta decay~\cite{bib:cuore1,bib:cuoricino,bib:cuore2}.
They are able to measure energies in the MeV region with a resolution of
the order of few keV.
One of the main sources of background in these searches is 
represented by the $\alpha$ particles emitted by natural radioactivity.
As predicted in~\cite{TabarellideFatis:2009zz} and 
demonstrated in~\cite{bib:cherenkov,bib:cherenkov_freddo},
the observation of light emitted by electrons in a {\TEO} bolometer
can provide a powerful tool to disentangle 
$\alpha$ from $\beta/\gamma$ radiation. 
According to~\cite{bib:cerenkov_noi},
the detected light is compatible with the \v Cerenkov emission 
and the absence of the contribution of scintillation light has been demonstrated 
in~\cite{IO}. 
The aims of the experiment presented in this paper are the evaluation of the number 
of \v Cerenkov photons emitted by the crystal and the study of the effect on the 
light collection of wrapping the crystal with a diffusive material.

\section{Experimental set-up}

\begin{figure}[h]
\begin{centering}
\includegraphics[width=10cm]{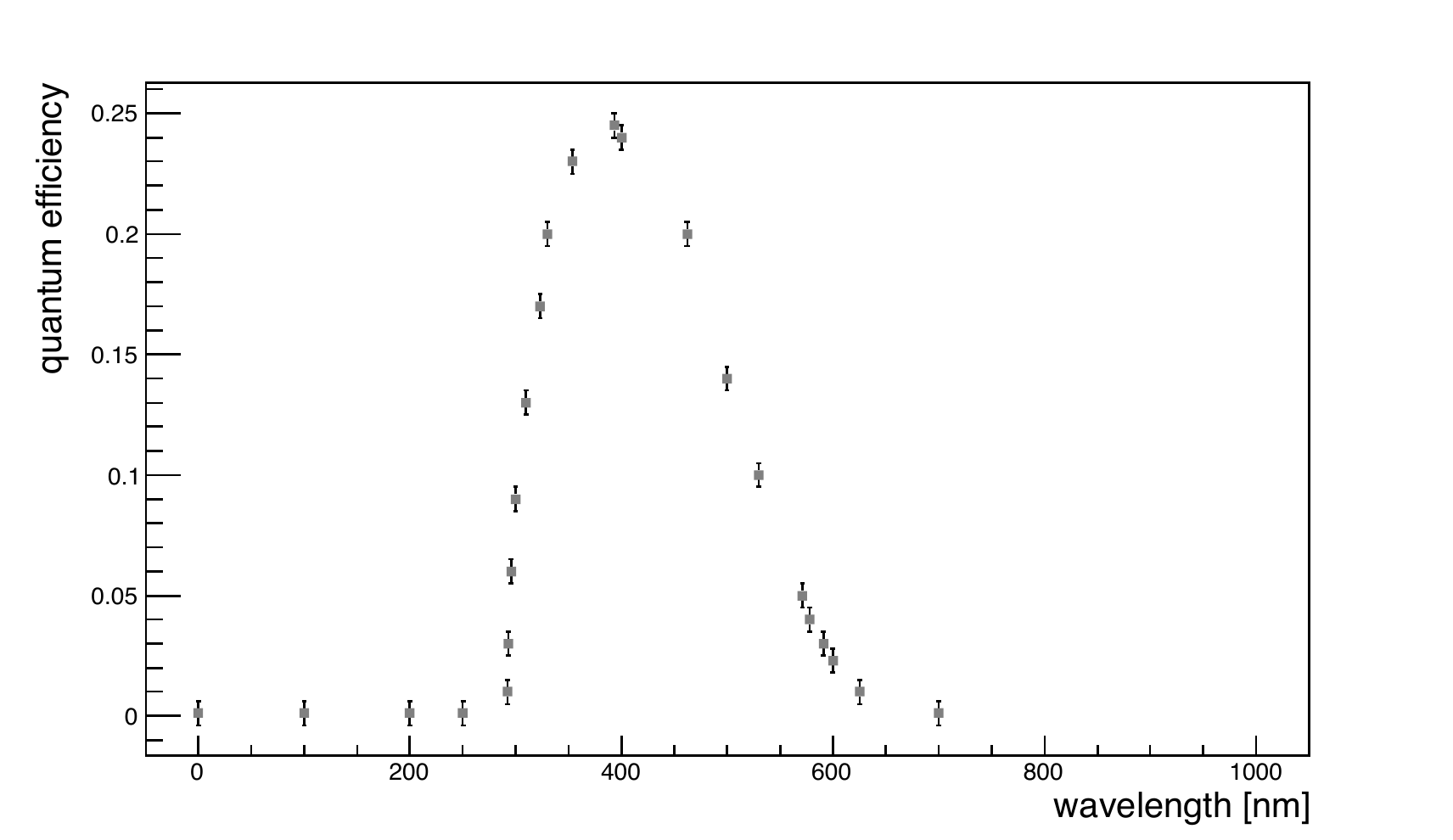}
\caption{Photo-cathode quantum efficiency of the PMTs 
(Hamamatsu R1924A data-sheet available at http://www.hamamatsu.com).}
\label{fig:qe}
\end{centering}
\end{figure}

A $5\times5\times5$~cm$^3$ crystal of {\TEO}
was placed in a light tight box.
In order to cross check the obtained results, 
the crystal was read out on two opposite faces by two
photo-multipliers (PMTs) (see Fig.\ref{fig:LYsetup}).  
Since the emitted light is expected to be due to 
\v Cerenkov photons, 
a good sensitivity in the ultraviolet region
is needed to maximize the light collection.
For this reason, two Hamamatsu R1924A PMTs were chosen. 
The 22~mm diameter window in borosilicate glass
provides these PMTs with a quantum efficiency larger than
10\% between 300~nm and 550~nm, with a peak of about 25\%
around 400~nm as shown in Fig.\ref{fig:qe}.
The PMTs were operated at 1200~V with a nominal gain of about 10$^7$.

The signals provided by the PMTs were acquired by
an oscilloscope with a bandwidth of 300~MHz and
a sampling frequency of 10 GS/s.

\section{Calibration of the photo-multipliers}

A first step, needed to measure the absolute light yield of the crystal,
is the calibration of the PMT response. In particular, the evaluation 
of the charge provided for a single photo-electron (p.e.) is required.
In order to perform this measurement, the light produced by a LED was sent
to the PMT by means of an optical fiber. A rectangular electric pulse, 60~ns wide, 
was used to drive the LED and to simultaneously trigger the oscilloscope acquisition.
A diaphragm placed upstream of the optical fiber
allowed to regulate the amount of light reaching the PMT window.

\subsection{Measurements with open diaphragm}
\label{open}
\begin{figure}
\begin{centering}
\includegraphics[width=10cm]{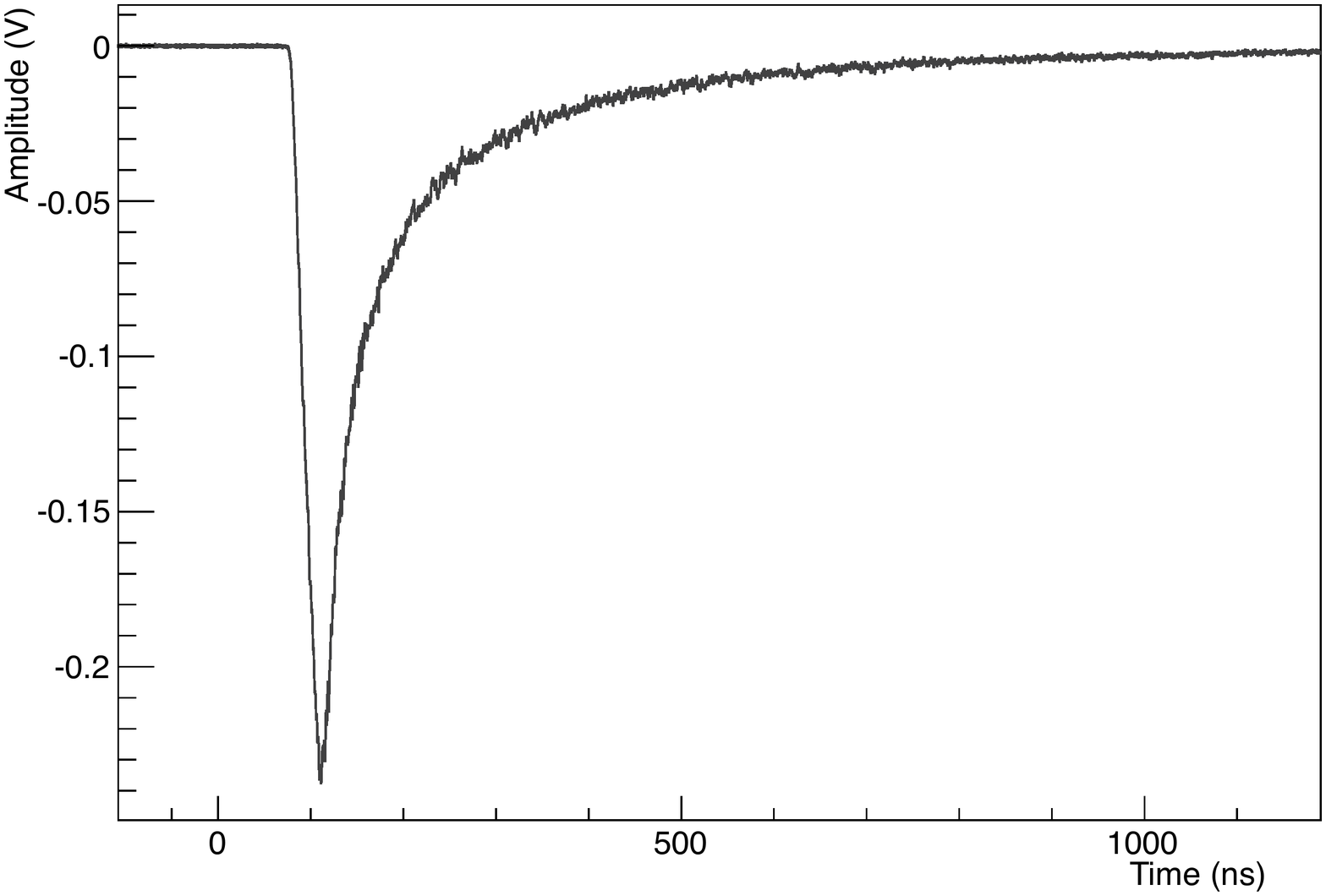} 
\caption{Average waveform of the PMT signal illuminated by the LED with a wide open diaphragm.}
\label{fig:dia6_av}
\end{centering}
\end{figure}

With the diaphragm wide open the average PMT waveform shown in Fig.~\ref{fig:dia6_av} is
obtained (the 0 of the time axis is arbitrary).

\begin{figure}
\begin{centering}
\includegraphics[width=10cm]{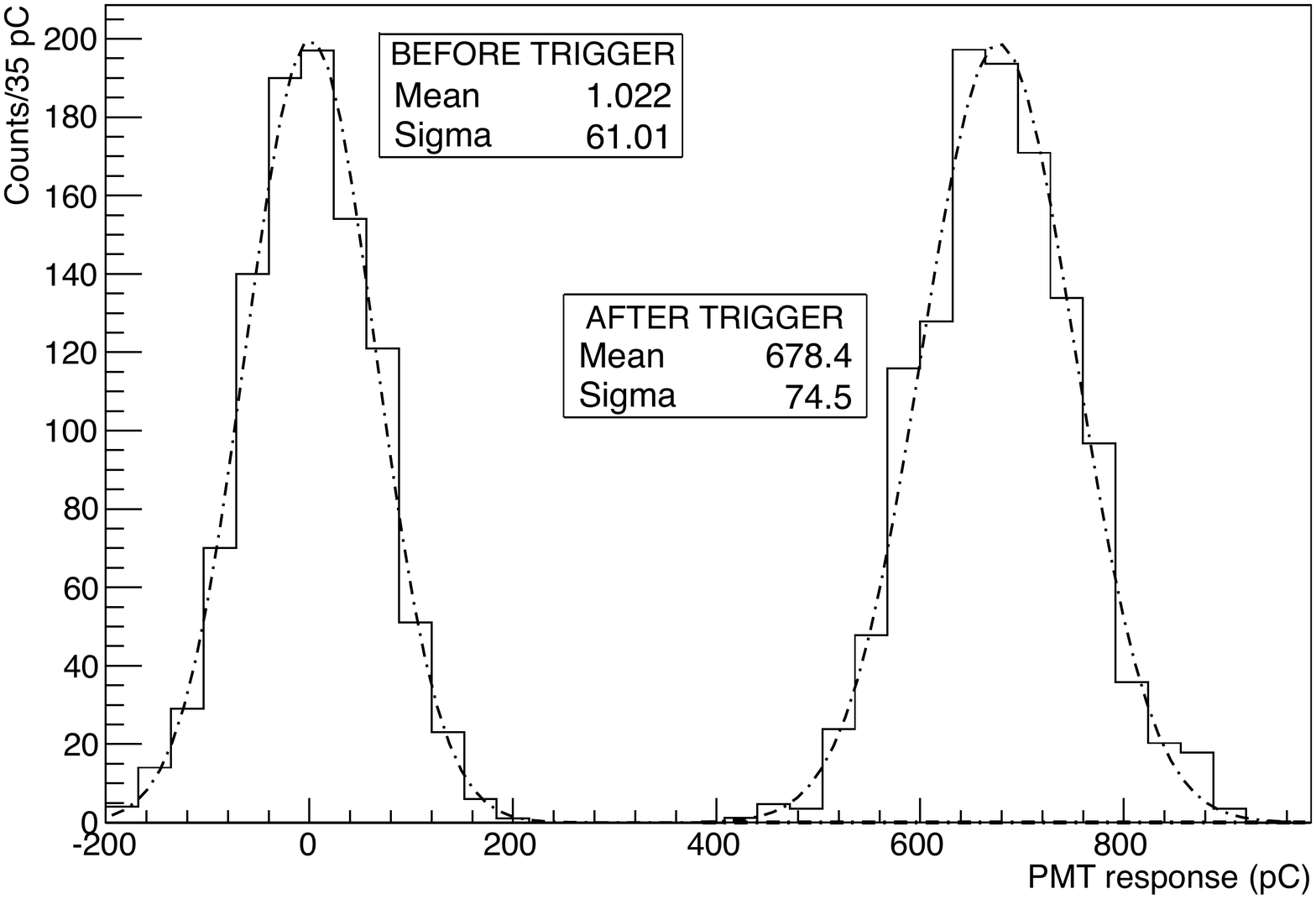} 
\caption{Charge spectrum of the pedestal (left) and of the signal (right),
with superimposed Gaussian fit,
obtained by illuminating the PMT with a blue LED with
a wide open diaphragm.}
\label{fig:dia6_spectra}
\end{centering}
\end{figure}

Fig.~\ref{fig:dia6_spectra} shows the charge spectrum obtained by
integrating event by event the charge produced by the PMT in a 1 $\mu$s time window before 
the trigger (left) and after the trigger (right). 

A Gaussian fit to the distributions allows to extract
the values of the fluctuations due to the electronics ($\sigma_{e}$)
and the total one ($\sigma_{t}$).
The value of the fluctuation 
of the charge signal ($\sigma_{s}$) can be obtained as:

$$
\sigma_{s} = \sqrt{\sigma^2_{t} - \sigma^2_{e}}
$$

and results to be $\sigma_{s} = 43 \pm 2$~pC.

In the hypothesis that $\sigma_{s}$
is mainly due to the statistical variation of the 
total number of photo-electrons ($n_{p.e.}$)
and that this number follows a Poisson distribution
the value of $n_{p.e.}$ can be calculated as:

$$
n_{p.e.} = \left(\frac{\mu}{\sigma_{stat}}\right)^2
$$ 

being $\mu = 687 \pm 1$~pC the average value of signal charge spectrum 
(Fig.~\ref{fig:dia6_spectra}, right).
It results that, in this configuration, $n_{p.e.}$ is~$242 \pm 30$, 
so that, the charge produced by 
each single p.e. ($\mu / n_{p.e.}$) is about $2.8 \pm 0.3$~pC.

\subsection{Measurements with almost closed diaphragm}
\label{closed}
\begin{figure}
\begin{centering}
\includegraphics[width=10cm]{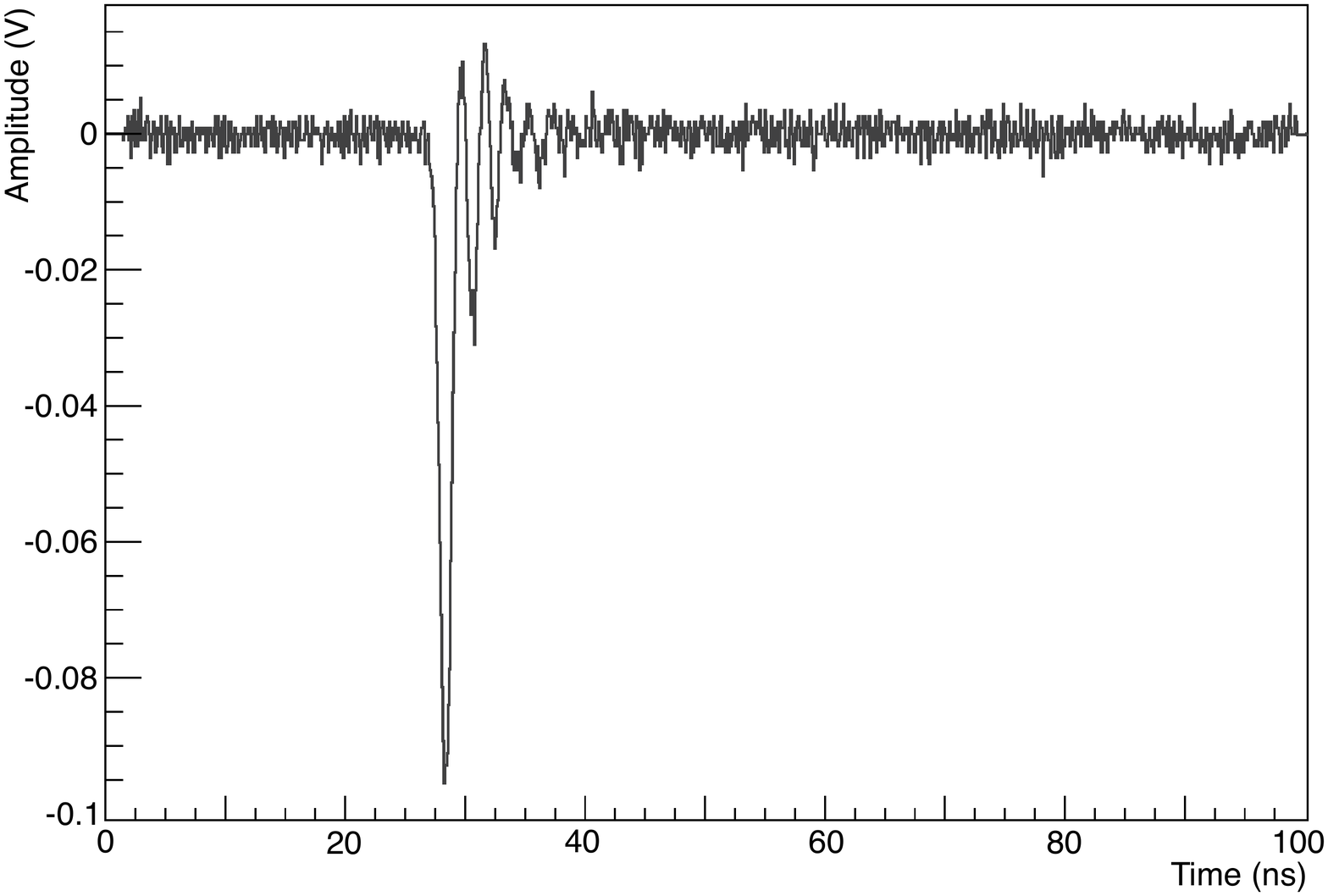} 
\caption{Example of a peak visible in the configuration with
almost closed diaphragm. Oscillations are due to a non perfect
impedance matching between the PMT and the oscilloscope.}
\label{fig:gInfPerst_closed}
\end{centering}
\end{figure}

To check the obtained results, a direct measurement of the charge 
produced by a single photo-electron was performed.
The diaphragm was almost completely closed in order to reduce
the amount of photons reaching the PMT.
In some events, one or, very rarely, more than one peaks as the 
one shown in Fig.\ref{fig:gInfPerst_closed} are visible.
Very likely these peaks are signals produced by a single photon
converted in an electron on the PMT photo-cathode.

In order to check this hypothesis the
distribution of the number of peaks per event is studied.
Each acquired waveform is analyzed and the number of times
the signal exceeds a threshold is counted. The threshold value is
optimized by studying the behavior of the noise before the trigger.
A time distance of at least 15~ns was required between two peaks
in order to avoid multiple counting of the same signal due to the
oscillations after the first peak (see Fig.~\ref{fig:gInfPerst_closed}).

\begin{figure}
\begin{centering}
\includegraphics[width=11cm]{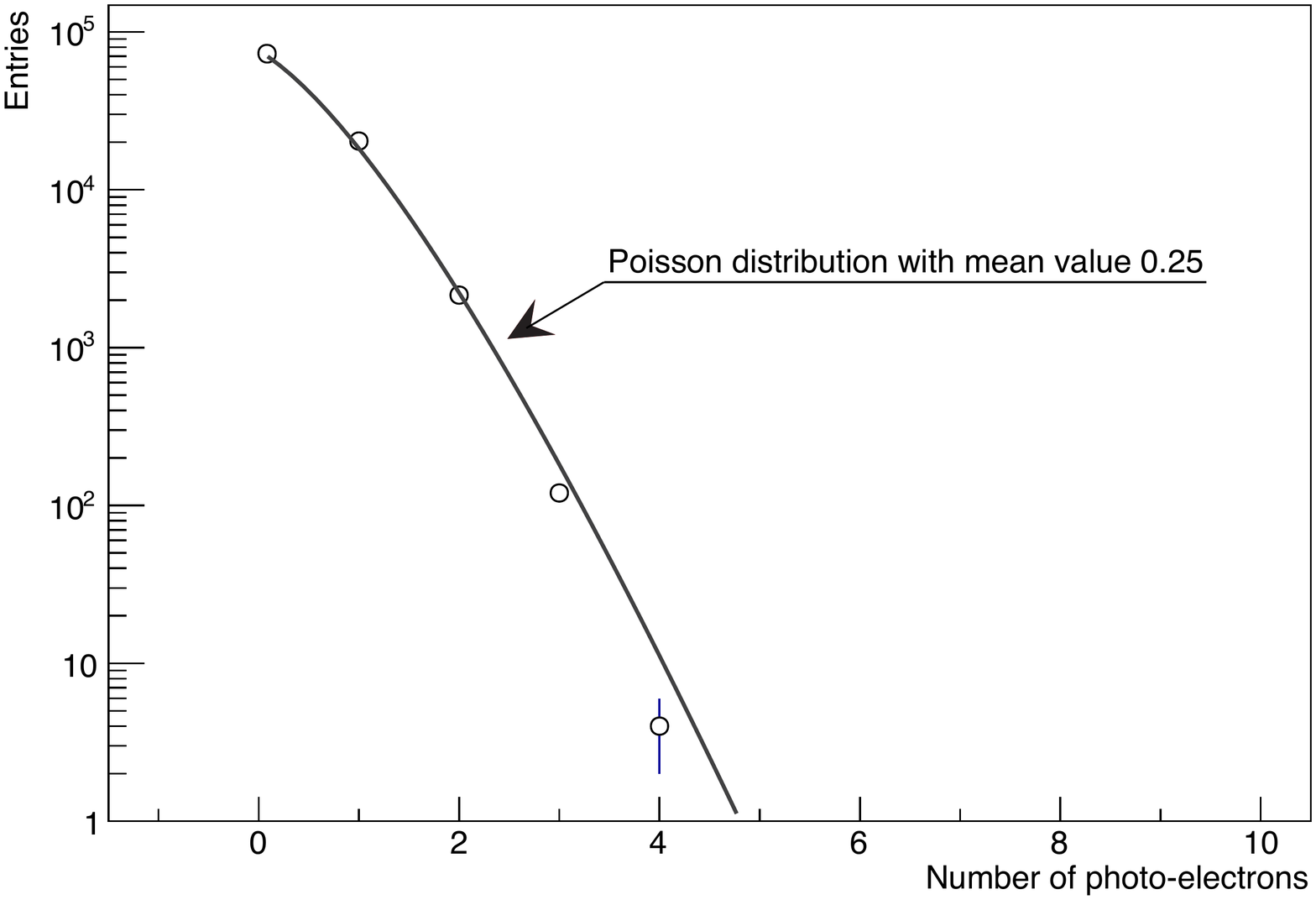}
\caption{Distribution of the number of peaks per event with 
superimposed a poissonian fit.}
\label{fig:poisson}
\end{centering}
\end{figure}

As it is shown in Fig.~\ref{fig:poisson} 
this quantity
follows a Poisson distribution as the number of photo-electrons is
expected to.

\begin{figure}
\begin{centering}
\includegraphics[width=10cm]{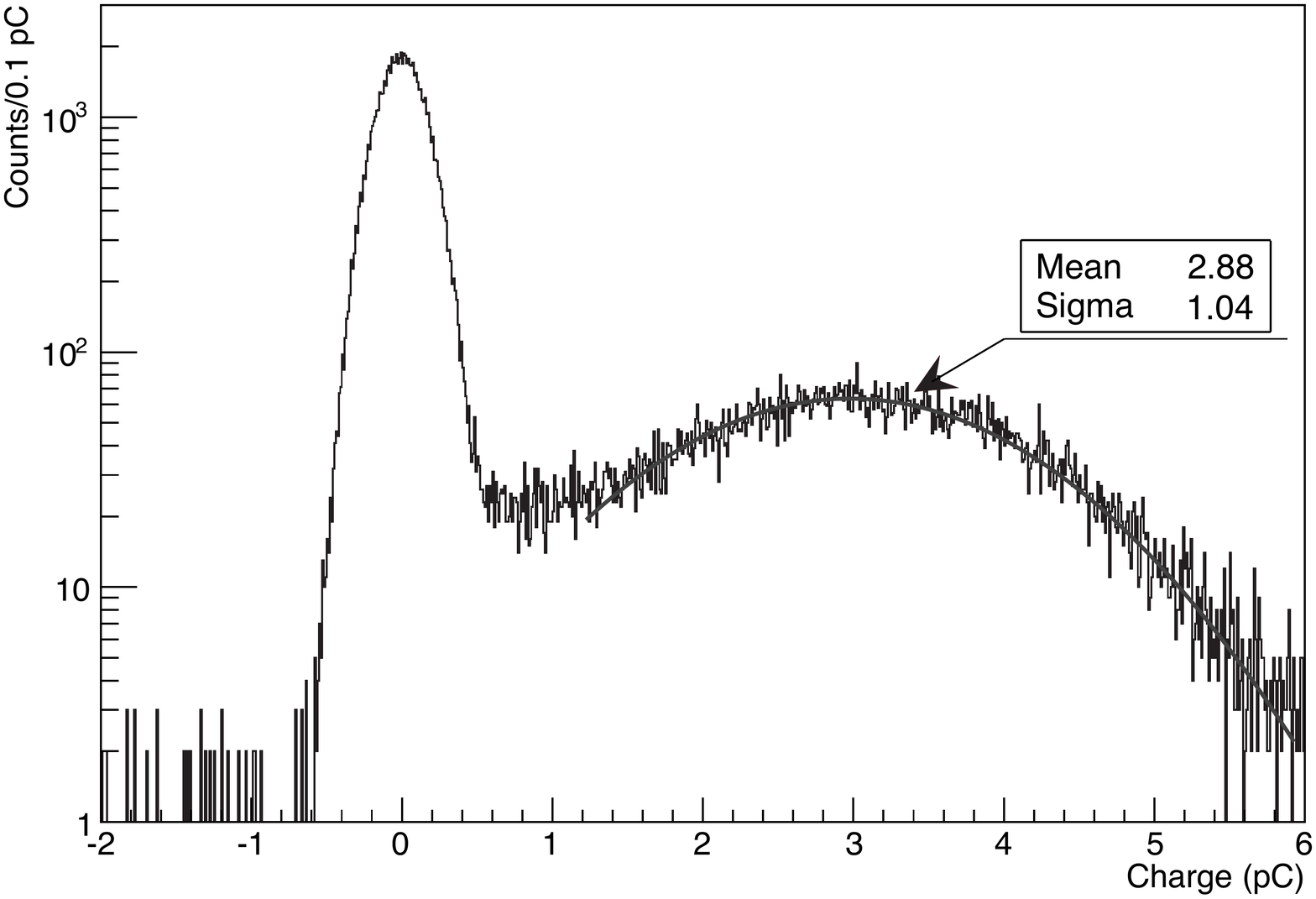}
\caption{Spectrum of the charge integrated in a 10~ns time gate around the peak 
in the configuration with almost closed diaphragm with
a superimposed Gaussian fit.}
\label{fig:pe_charge}
\end{centering}
\end{figure}
Moreover, a Gaussian fit to the distribution of the charge integrated 
in a 10~ns time gate around the peak
(2~ns before and 8~ns after the maximum amplitude)
returns a value of $2.88 \pm 1.04$ pC
(Fig.~\ref{fig:pe_charge}) that 
is in very good agreement with the one found in \ref{open}.
It is therefore possible to conclude that these peaks
are signals due to single photo-electrons and to confirm that
2.88 pC is the corresponding charge.
The same calibration performed on the opposite PMT gives a value of
about 2.70 pC.

\section{Light yield measurement}
\begin{figure}
\begin{centering}
\includegraphics[width=12cm]{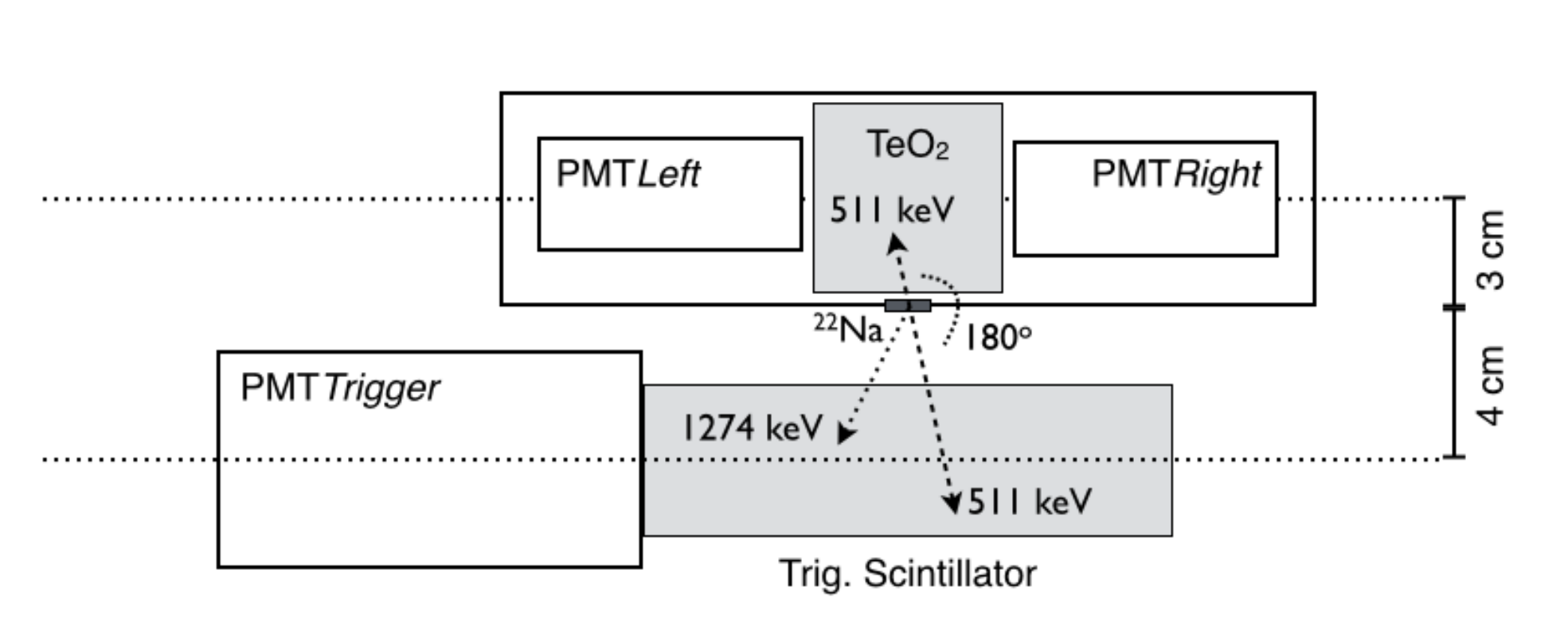}
\caption{Experimental set-up.}
\label{fig:LYsetup}
\end{centering}
\end{figure}

In order to measure the absolute light yield of the crystal the set-up shown in 
Fig.~\ref{fig:LYsetup} was prepared.
The idea is to measure the number of photo-electrons
provided by the PMT
for a given value of energy released in the crystal.

A $^{22}$Na source was placed on one side of the black box, 
exposed towards a plastic scintillator. The signal of the scintillator
PMT was sent to the oscilloscope, used as trigger and acquired.
The source in use
emits two back-to-back 511 keV photons, and, simultaneously,
a 1274 keV photon, uncorrelated in space.
Given the geometry of the system, in the triggered events 
one of the two 511 keV photons
always reaches the {\TEO} crystal.
In order to select events where
only the 511 keV photon reaches the crystal
the threshold on the trigger signal
was tuned to select the events where 
the 1274 keV photon hits the trigger scintillator. 

From the comparison of the results obtained in runs with and without
the $^{22}$Na source it was evaluated that about 10\% of triggers
were due to noise in the trigger PMT.

\begin{figure}
\begin{centering}
\includegraphics[width=7cm, angle=90]{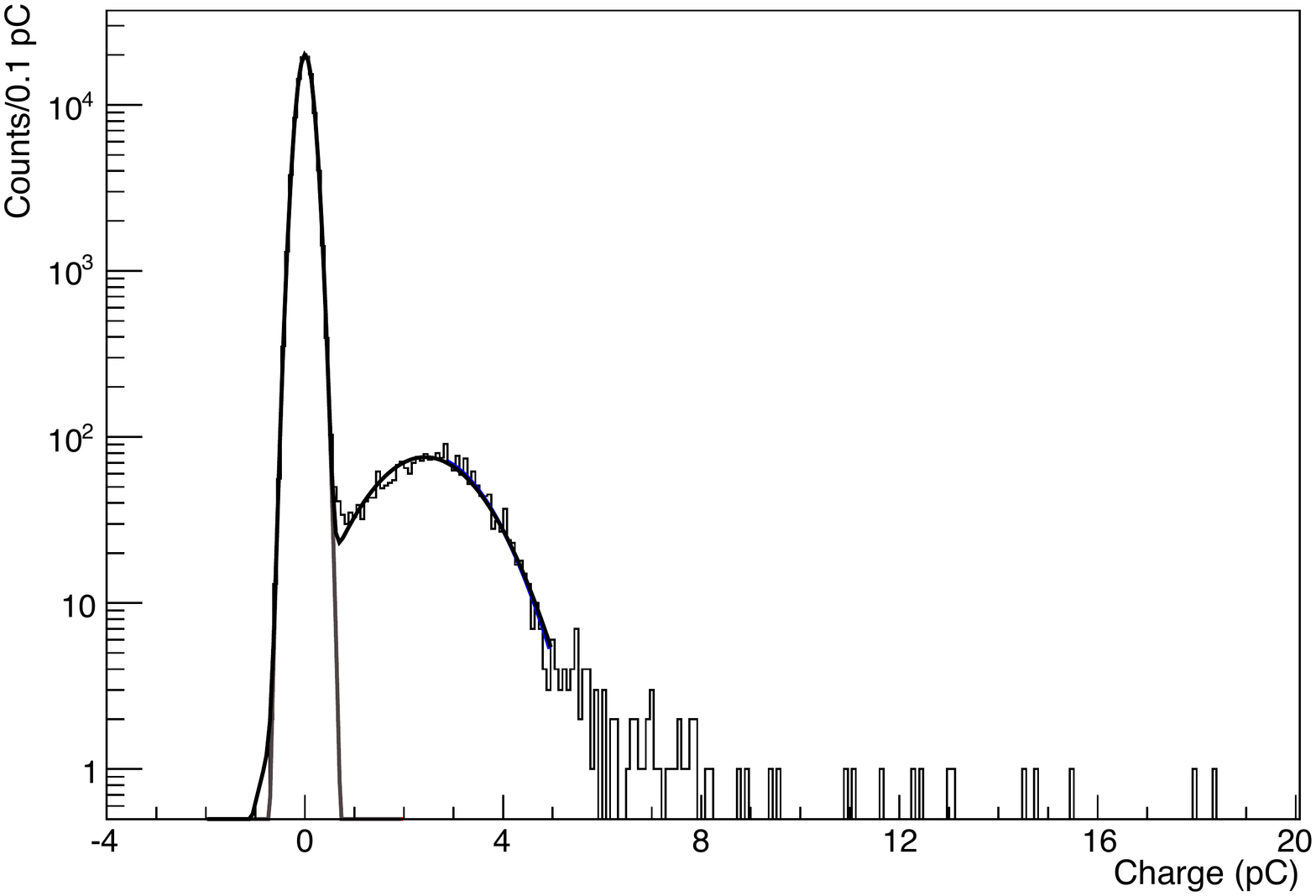}
\caption{Example of a charge spectrum of TeO$_2$ crystal
obtained with $^{22}$Na source. The two visible Gaussian fits
were used to determine the number of events with 0 and 1 p.e.}
\label{fig:teo}
\end{centering}
\end{figure}

The response spectrum of the {\TEO} crystal in a standard run 
is represented in Fig.~\ref{fig:teo}
with a superimposed double Gaussian fit.

It shows a narrow peak around 0, a large peak around 1 photo-electron and a 
small tail for higher values.
Since the number of produced photo-electrons is expected to follow a Poisson distribution,
if P($k$) is the probability of detecting $k$ photo-electrons, the average number $\mu$ 
of collected photo-electrons is given by:

\begin{equation}
\mu = \frac{P(1)}{P(0)}
\label{form:mu}
\end{equation}

where P(0) and P(1) are evaluated from the results of the fit, 
after subtracting the noise contribution.

The values of $\mu$ were calculated for the two PMTs and
were found to be:
$\mu_{L} = 0.015 \pm 0.003$ and $\mu_{R} = 0.016 \pm 0.003$.

\section{Monte Carlo simulation}

\begin{figure}[h]
\begin{centering}
\includegraphics[width=10cm]{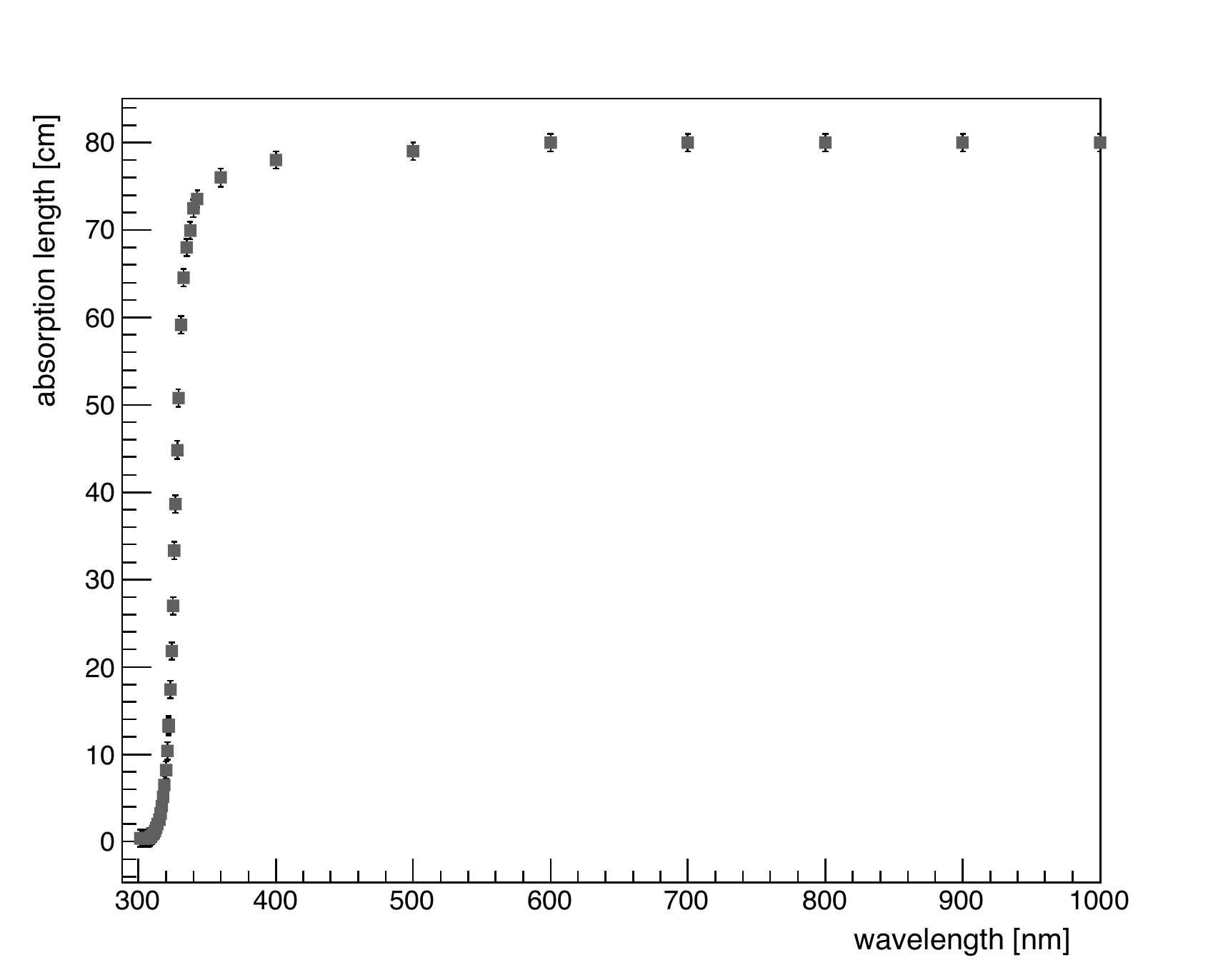}
\caption{Result of the measurement of the 
absorption length of a {\TEO} crystal.}
\label{fig:optprop}
\end{centering}
\end{figure}

A Monte Carlo simulation of the \v{C}erenkov radiation produced by the 511~keV photon 
interactions with the {\TEO} crystal and their propagation in the experimental 
set-up was performed. The simulation is divided in two parts:
\begin{itemize}
\item in the first part the total amount of \v{C}erenkov photons for each 
interaction is evaluated;
\item in the second part the propagation of the \v{C}erenkov photons inside 
the different components of the experimental set-up is reproduced by means 
of Litrani~\cite{litrani}, a software developed to simulate the propagation 
of the optical photons in any type of optical media and able to model 
the response of the photo-multipliers.
\end{itemize}
The optical properties required by Litrani to reproduce the propagation 
of the optical photons within the {\TEO} crystal are
the ordinary and extraordinary refractive index and 
the absorption length.
While the refractive indices were taken from~\cite{bib:index},
the absorption length was measured at room temperature
and the result is shown in Fig.~\ref{fig:optprop}.

The light diffusion due to the surface properties 
of the crystal is taken into account in the simulation.
The two faces in front of the PMTs were simulated to have
a high polishing quality, close to optical polishing grade,
while the other four were simulated with higher roughness.

The photo-cathode quantum efficiency (Fig.~\ref{fig:qe})
and the refractive index of the borosilicate window of the PMT
are taken into account by Litrani.

\begin{figure}[h]
\begin{centering}
\includegraphics[width=7cm, angle=90]{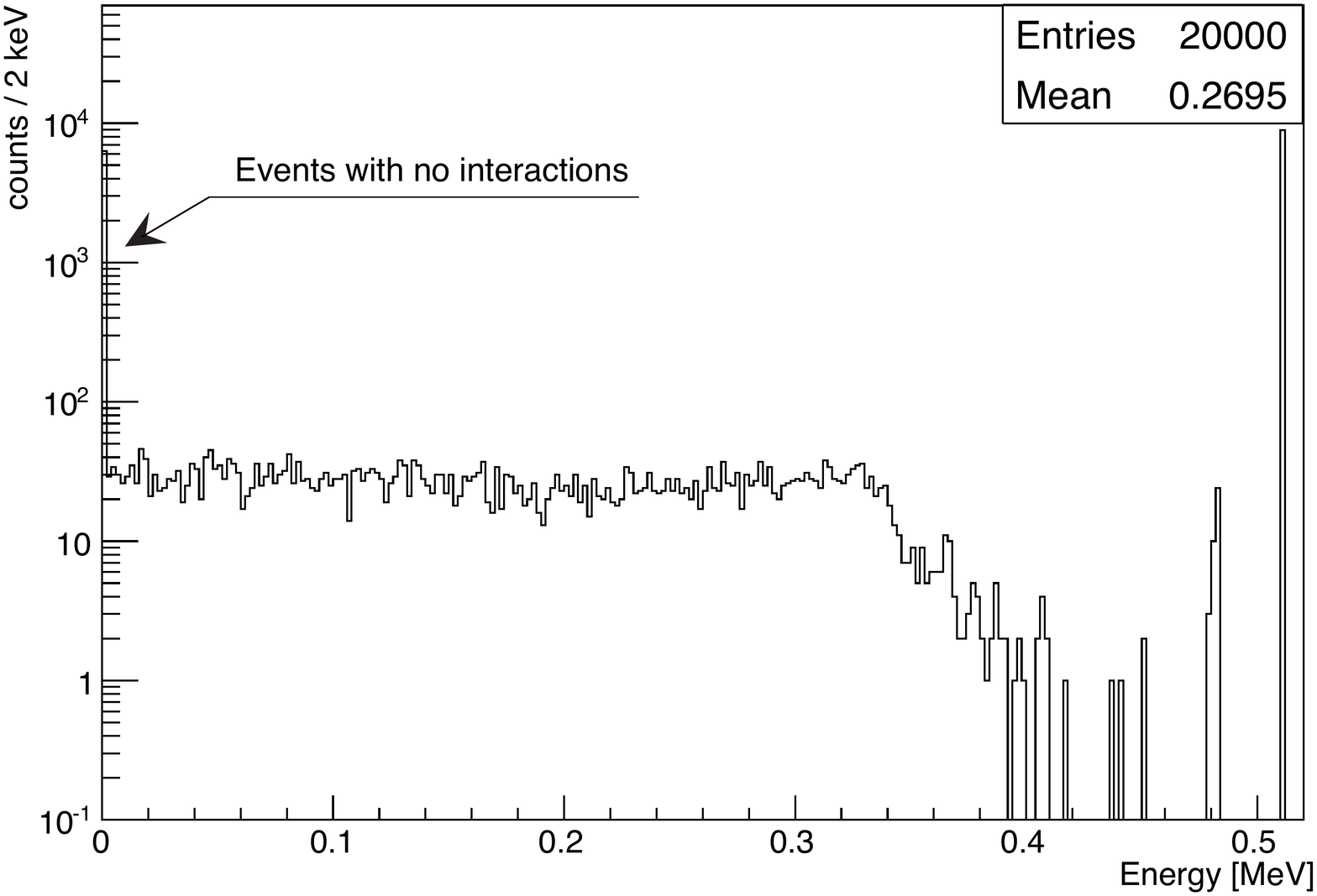}
\caption{Spectrum of the energy deposited by the 511 keV photons in the 
{\TEO} crystal as obtained from the Monte Carlo simulation.}
\label{fig:enedep}
\end{centering}
\end{figure}

The simulation starts with the emission of $2\cdot10^{4}$~photons 
with an energy of 511~keV,
from a point-like source located at 5~mm from the {\TEO} crystal with an 
isotropic angular distribution inside the solid angle covered by the trigger 
scintillator (see Fig.~\ref{fig:LYsetup}). The energy deposition in the {\TEO} crystal 
for each interaction is shown in Fig.~\ref{fig:enedep}.

\begin{figure}[h]
\begin{centering}
\includegraphics[width=10cm]{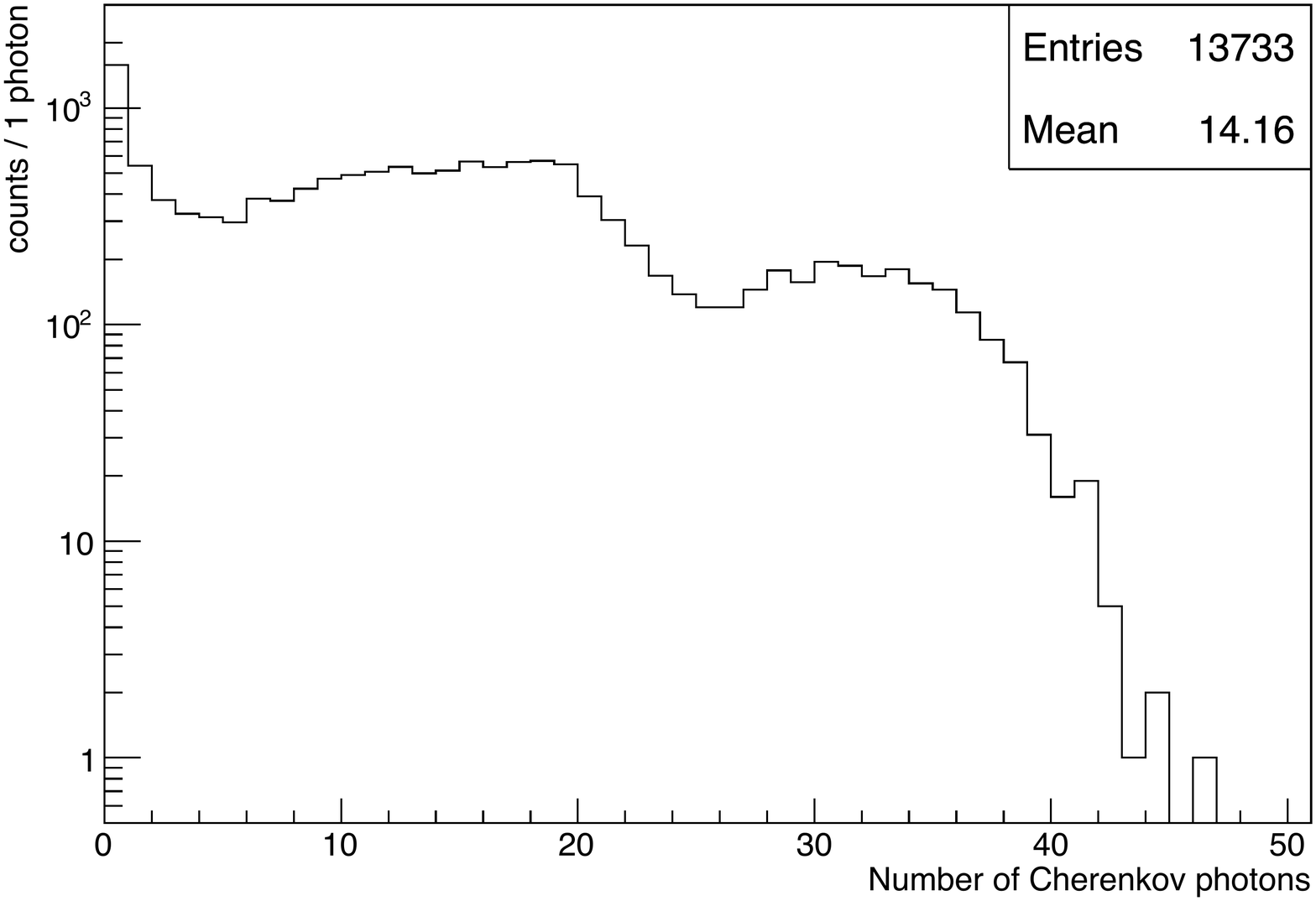}
\caption{Distribution of the number
of \v{C}erenkov photons produced 
per interaction within the {\TEO} crystal
as obtained from the Monte Carlo simulation.}
\label{fig:wdistro}
\end{centering}
\end{figure}

According to the simulation, 
given the geometry, in about 31\% 
of triggered events the 511~keV photon crossing the {\TEO}
crystal does not interact with it, and the
average energy released within the crystal
is about 0.269~MeV.
For each interaction within the crystal, the total 
number of \v{C}erenkov photons 
with a wavelength in the range $300 \div 1000$~nm
is evaluated and its distribution is shown in Fig.~\ref{fig:wdistro}.

Since the \v{C}erenkov threshold for an electron in {\TEO} crystal 
is about 50 keV \cite{TabarellideFatis:2009zz} 
the probability of having zero emitted \v{C}erenkov photons
was found to be about 12.1\%. 

The peak around 30 photons is due to photo-electric effect
within the crystal with a release of the whole energy (511 keV) within it.
The shoulder between 10 and 20 photons is instead 
due to single or multiple Compton scattering.

From the ratio between the average number of produced photons
and the mean energy released, a value of about 52~\v{C}erenkov
photons per MeV released can be calculated.

By means of Litrani the photons are thus simulated inside the experimental set-up. 
The number of photons detected by the PMTs was found to be $\mu$=$0.015\pm0.002$ 
in good agreement with the measured one
representing a good validation 
of the adopted simulation method.

\section{Light yield optimization}
\subsection{Experimental measurements}

\begin{figure}
\begin{centering}
\includegraphics[width=7cm, angle=90]{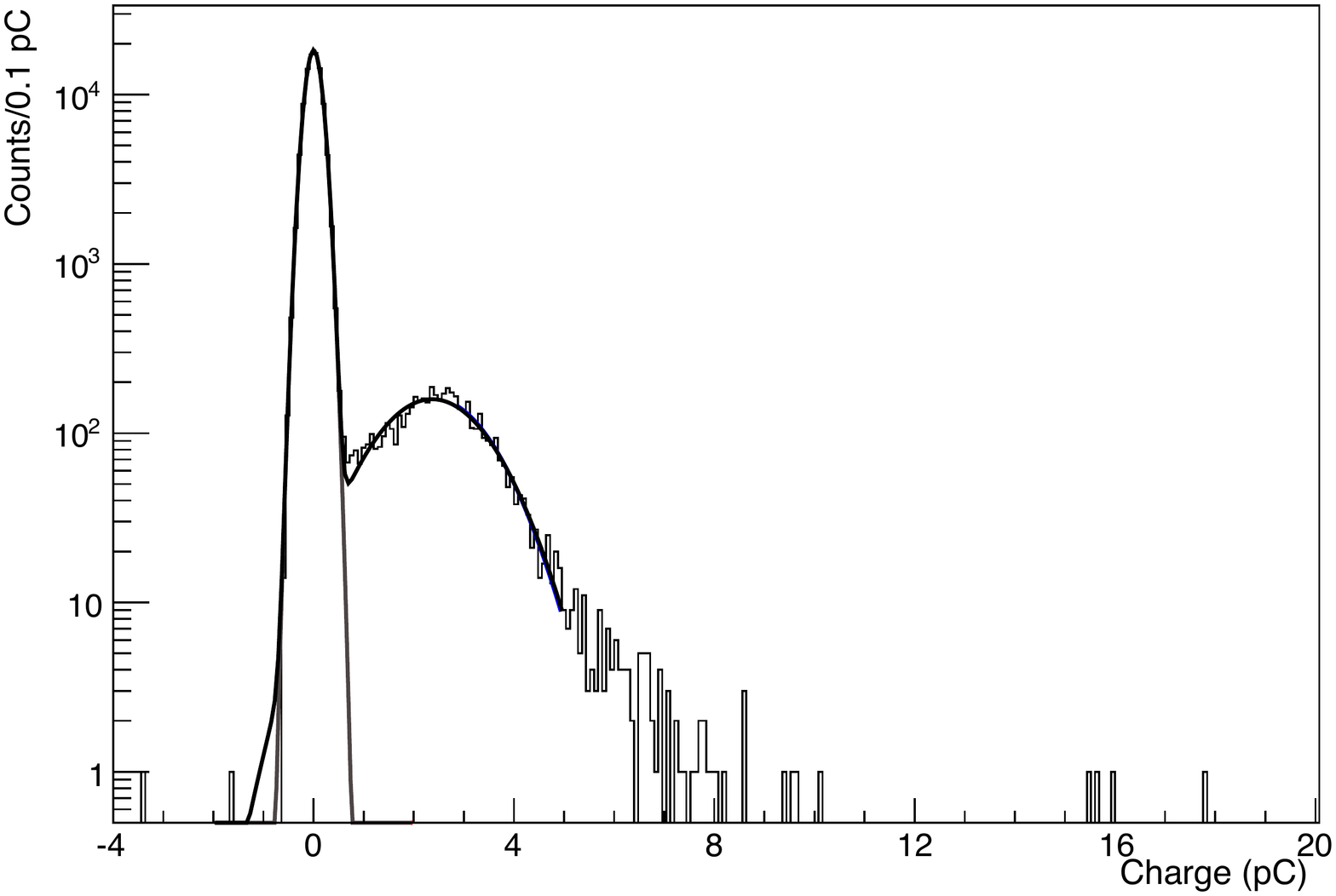}
\caption{Charge spectrum obtained with $^{22}$Na source
with a {\TEO} crystal fully wrapped with PTFE.}
\label{fig:wrapping_spectra}
\end{centering}
\end{figure}

In order to maximize the amount
of light collection in the PMTs,
all the crystal faces were wrapped with Polytetrafluoroethylene (PTFE),
a white and diffusive material.
The measured response spectrum is shown in 
Fig.~\ref{fig:wrapping_spectra}.

By means of the formula~\ref{form:mu}
the average number of photo-electrons
were computed and found to be
$\mu_{L}$=$0.036\pm0.002$ and $\mu_{R}$=$0.036\pm0.002$.

The use of the PTFE on all faces of the crystal
is able to increase by a factor 2.4 
the number of photo-electrons detected.

\begin{figure}[h]
\begin{centering}
\begin{tabular}{cc}
\includegraphics[width=8cm]{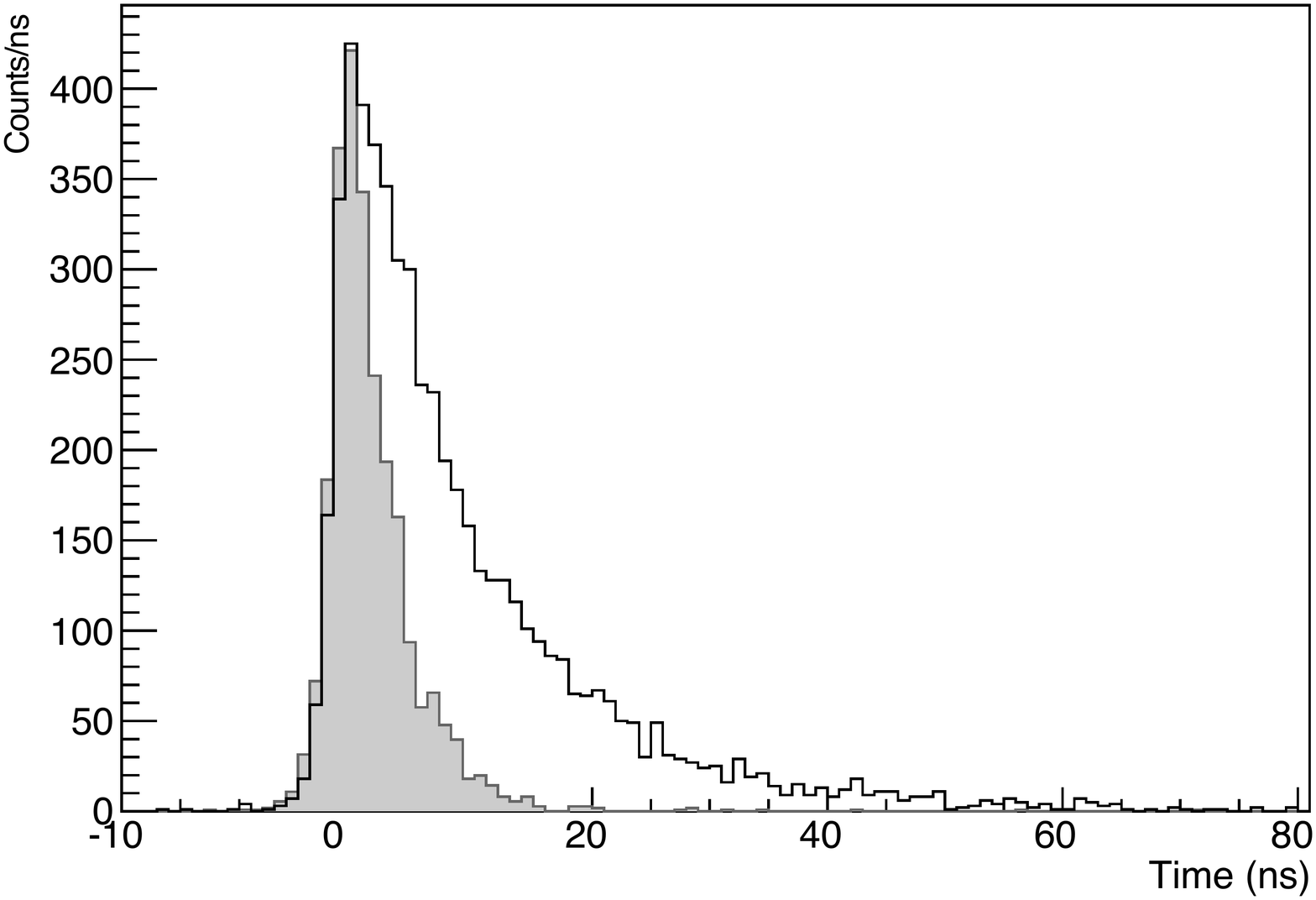} 
\end{tabular}
\caption{Measured distributions of the arrival time of the \v{C}erenkov photons
on the PMT without (grey) and with (white) wrapping on the crystal.}
\label{fig:hjmax}
\end{centering}
\end{figure}

An important aspect is that
the signal arrival time distribution (Fig.~\ref{fig:hjmax})
shows a large tail on high values when the crystal is wrapped.
This confirms that the light yield increase is due to
photons that are reflected and diffused by the wrapping
and travel up to $30\div40$~ns before reaching the PMT
instead of exiting the crystal from the lateral faces.

\subsection{Monte Carlo simulation}

The effects of the PTFE wrapping were also studied by means of dedicated simulations.
The number of photoelectrons was found to be:
$\mu$=$0.034\pm0.002$ in very good agreement with the experimental data.

\begin{figure}[h]
\begin{centering}
\includegraphics[width=8cm]{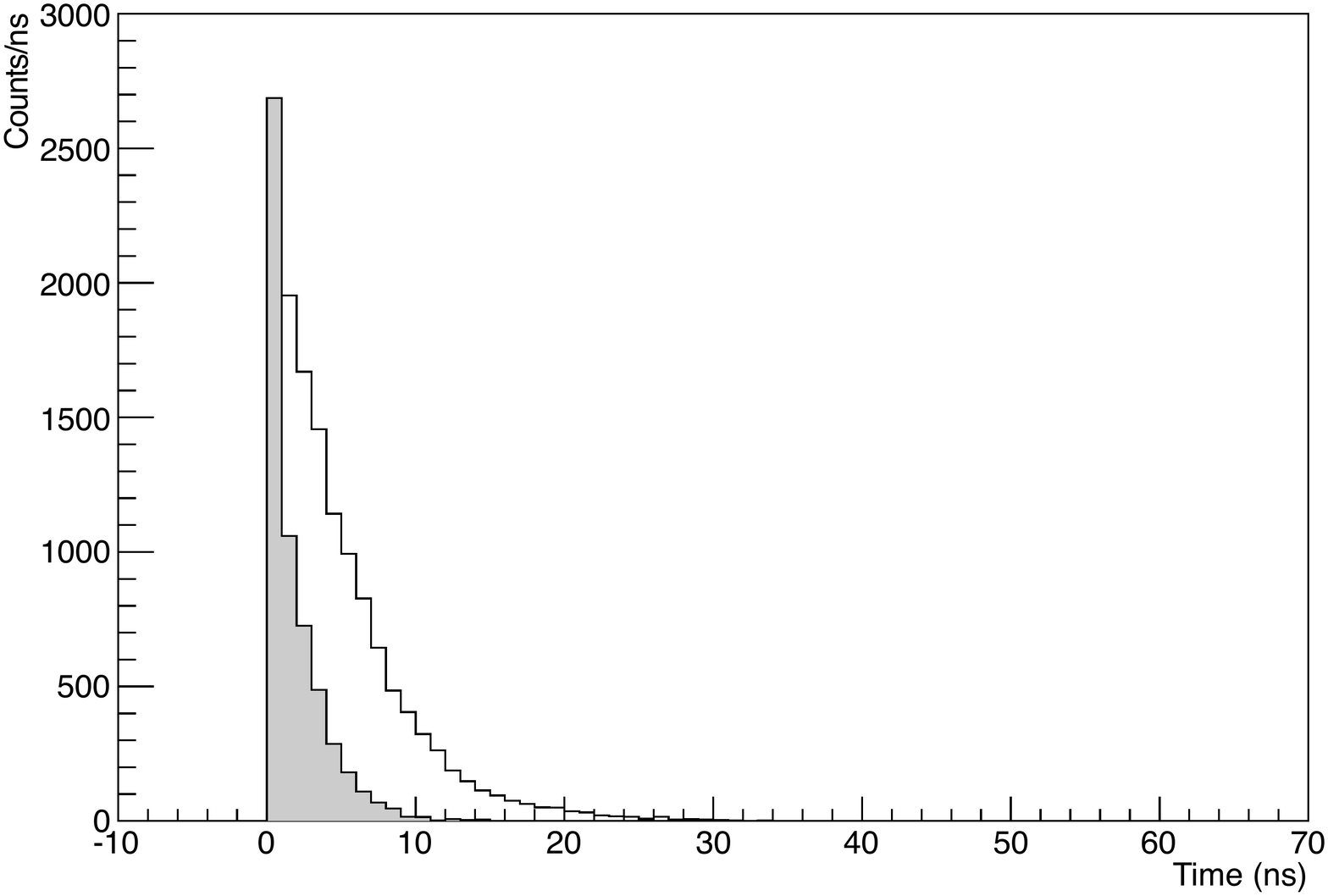}
\caption{Distributions of the arrival time of the \v{C}erenkov photons
on the PMT without (grey) and with (white) wrapping on the crystal as 
resulted in the simulation.}
\label{fig:MC_timeCompare}
\end{centering}
\end{figure}

Also the simulation results show (Fig.~\ref{fig:MC_timeCompare})
that, thanks to the reflection on the wrapping,
several photons that would have been lost
are instead driven toward the PMTs and are detected
after having traveled for up to 30~ns.

The smaller tail for large times that is 
found in the simulation can be explained
with the time jitter of the PMT that was 
not taken into account in the simulation.

\section{Conclusion}

The amount of light produced by a TeO$_2$ crystal exposed
to the 511 keV photons produced by a $^{22}$Na radioactive source
was measured by means of two PMTs.
Without any wrapping, an average value of 0.015$\div$0.016 
photo-electrons was measured (i.e. about 0.055 p.e./MeV).
By means of a Monte Carlo simulation, it was possible
to derive that the number of primary 
\v{C}erenkov photons produced per MeV
released in the crystal is about 52. 
The effect of a reflective and diffusive wrapping
was studied.
The most promising result was obtained by covering all lateral faces 
with PTFE, that allowed to increase the number
of photons reaching the PMTs by a factor 2.4.


\begin{thebibliography}{9}
\bibitem{bib:cuore1}
C. Arnaboldi {\it et al.}, 
``CUORE: a cryogenic underground observatory for rare events,''
Nucl.\ Instrum.\ Meth.\ A {\bf 518} 775 (2004). 

\bibitem{bib:cuoricino}
E. Andreotti {\it et al.}, 
``$^{130}$Te neutrinoless double-beta decay with CUORICINO,''
Astropart.\ Phys.\ {\bf 34} 822 (2011).

\bibitem{bib:cuore2}
D. Artusa, {\it et al.},  (CUORE Collaboration), 
``Searching for neutrinoless double-beta decay of $^{130}$Te with CUORE'', submitted (2014).

\bibitem{TabarellideFatis:2009zz}
 T.~Tabarelli de Fatis,
 ``Cherenkov emission as a positive tag of double beta decays in bolometric experiments,''
 Eur.\ Phys.\ J.\ C {\bf 65} (2010) 359.

\bibitem{bib:cherenkov}
J.~W.~Beeman{\it et al.},
  ``Discrimination of alpha and beta/gamma interactions in a TeO$_2$ bolometer,''
  Astropart.\ Phys.\  {\bf 35}, 558 (2012).

\bibitem{bib:cherenkov_freddo}
N.~Casali, {\it et al.},``TeO$_2$ bolometers with Cherenkov signal tagging: 
towards next-generation neutrinoless double beta decay experiments,''
arXiv:1403.5528 [physics.ins-det].

\bibitem{bib:cerenkov_noi}
  F.~Bellini {\it et al.},
  ``Measurements of the Cherenkov light emitted by a TeO2 crystal,''
  JINST {\bf 7} (2012) P11014.

\bibitem{IO} N.~Casali {\it et al.}, 
``Monte Carlo simulation of the Cherenkov radiation emitted by {\TEO} crystal when crossed by cosmic muons'' 
  Nucl.\ Instrum.\ Meth.\ A 
  http://dx.doi.org/10.1016/j.nima.2013.07.024.

\bibitem{bib:geant4}
  S.~Agostinelli {\it et al.},
  Nucl.\ Instrum.\ Meth.\ A {\bf 506}, (2003) 250.

\bibitem{bib:geant4_2}
  J.~Allison {\it et al.},
  IEEE Transactions on Nuclear Science 53 No. 1 (2006) 270-278. 

\bibitem{litrani} 
Fran\c{c}ois Xavier Gentit, 
``Litrani: A general purpose Monte Carlo program simulating light propagation in isotropic or anisotropic media'' 
  Nucl.\ Instrum.\ Meth.\ A {\bf 486} (2002) 35.

\bibitem{bib:index}
N. Uchida, ``Optical properties of single-crystal paratellurite (TeO$_2$)'', Phys. Rev.B 4, 3736-3745 (1971).

\end{thebibliography}
\end{document}